 \documentclass[preprint,review,12pt]{elsarticle}

\usepackage{amsmath,amssymb,amsfonts}
\usepackage[ruled, linesnumbered]{algorithm2e}
\usepackage{makecell}
\usepackage{hyperref}
\newcommand{\orcid}[1]{\href{https://orcid.org/#1}{\includegraphics[width=10pt]{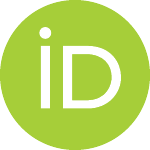}}}

\journal{Information Sciences}

\usepackage{xcolor}

\def\methodName{RelJoin}

\begin{document}

\begin{frontmatter}


\title{\methodName: Relative-cost-based Selection of
\\ Distributed Join Methods for Query Plan Optimization}


\author[smbu,eipc]{Feng Liang\orcid{0000-0002-8542-9871}\corref{cor1}}
\ead{fliang@smbu.edu.cn}
\author[hku]{Francis C.M. Lau}
\author[hku]{Heming Cui}
\author[hkbu]{Yupeng Li\orcid{0000-0001-9652-3321}}
\author[fnu,klqm,pku]{Bing Lin\orcid{0000-0001-5874-4748}}
\author[smbu,eipc]{Chengming Li}
\author[smbu,eipc,bit]{Xiping Hu\orcid{0000-0002-4952-699X}\corref{cor1}}
\ead{huxp@smbu.edu.cn}
\cortext[cor1]{Corresponding authors}

\affiliation[smbu]{organization={Artificial Intelligence Research Institute, Shenzhen MSU-BIT University},
            city={Shenzhen},
            country={China}}
\affiliation[eipc]{organization={Guangdong-Hong Kong-Macao Joint Laboratory for Emotional Intelligence and Pervasive Computing, Shenzhen MSU-BIT University},
            city={Shenzhen},
            country={China}}
\affiliation[bit]{organization={School of Medical Technology, Beijing Institute of Technology},
            city={Beijing},
            country={China}}

\affiliation[hku]{organization={Department of Computer Science, The University of Hong Kong},
            city={Hong Kong},
            country={China}}
            
\affiliation[hkbu]{organization={Department of Interactive Media, Hong Kong Baptist University},
            city={Hong Kong},
            country={China}}
            
            
\affiliation[fnu]{organization={College of Physics and Energy, Fujian Normal University},
            city={Fuzhou},
            country={China}}
\affiliation[klqm]{organization={Key Laboratory of Quantum Manipulation and New Energy Materials},
           city={Fuzhou},
           country={China}}
\affiliation[pku]{organization={School of Computer Science, Peking University},
            city={Bejing},
            country={China}}

\begin{abstract}
	Selecting appropriate distributed join methods for logical join operations in a query plan
	is crucial for the performance of data-intensive scalable computing (DISC).
	Different network communication patterns in the data exchange phase
	generate varying network communication workloads and
	significantly affect the distributed join performance.
	However, most cost-based query optimizers focus on the local computing cost 
	and do not precisely model the network communication cost. 
	We propose a cost model for various distributed join methods
	to optimize join queries in DISC platforms.
	Our method precisely measures the network and local computing workloads in different execution phases, 
	using information on the size and cardinality statistics of datasets and cluster join parallelism.
	Our cost model reveals the importance of the relative size of the joining datasets.
	We implement an efficient distributed join selection strategy, known as \methodName\
	in SparkSQL, which is an industry-prevalent distributed data processing framework. 
	\methodName\ uses runtime adaptive statistics for accurate cost estimation
	and selects optimal distributed join methods
	for logical joins to optimize the physical query plan.
	The evaluation results on the TPC-DS benchmark 
	show that \methodName\ performs best in 62 of the 97 queries
	and can reduce the average query time by 21\% 
	compared with other strategies.\footnote{The
		source code and evaluation data are available 
		at: https://github.com/liangfengsid/relJoin.}
	
\end{abstract}




\end{frontmatter}



\section{Introduction}
Join query optimization is crucial for improving database query performance. 
Cost-based query plan optimization is an intensively studied direction 
in join query optimization~\cite{costOracle06,costSpark18, queryOpt15,queryOptGood15,costSparkSqlJoin18,forecastCost15,costAdaptive09, costOptAI14,predictTime13} 
and is adopted by almost all mainstream database 
systems such as PostgreSQL~\cite{predictTime13} and SparkSQL~\cite{sparkSql15}.
Cost-based optimization usually estimates the disk I/O, CPU, and memory costs of 
several query plan candidates based on the statistics of the datasets
and selects the one with the lowest cost estimate.


Although distributed join queries can use many traditional cost-based 
query plan optimization strategies that are designed for a single computer node, 
they face unique challenges owing to their distributed nature. 
First, physical methods for distributed joins are more complicated than those of local joins.
A distributed join involves a data exchange phase~\cite{sparkSql15,costSpark18}
in which data are transferred to different partitions via the communication
network before the local join operation is executed in each node.
Instead of planning a sort join or hash join for local joins, 
the query planner for distributed joins has other options, including
the broadcast hash join, shuffle hash join
and shuffle sort join. Different physical join methods 
may perform completely differently.
Second, the network communication cost, or network cost,  
is a non-negligible part of the total cost, which complicates the cost model.
Broadcasting a complete replicate of a table for all nodes in a large cluster 
or shuffling the data of a large table inevitably incurs considerable network
traffic.
This network cost may be substantial for large datasets or clusters,
which can result in a performance bottleneck for distributed operations that 
require data repartitioning~\cite{confluence17,broadcastJoin14}.
The cost model must also properly assign weights to the network and other resource costs.
Third, the consequences of using biased dataset statistics for cost estimation can be amplified 
in distributed joins. For example, if the size-related statistics of a large table 
are mistakenly estimated as small, the query optimizer may select the incorrect join order or method.
An inferior choice between the sort and hash methods 
may result in little performance difference in local joins~\cite{sortVsHash13},
whereas broadcasting a large table instead of shuffling it can be disastrous for distributed joins~\cite{adaptiveBig13}.

Therefore, a precise cost model is imperative to select the best physical join 
method for distributed joins. 
Several studies on cost-based optimization (CBO) have focused on
logical join order optimization~\cite{joinOrder09,queryOptGood15,costAdaptive09,randomJoinOrder97,joinOrder03} because local joins can benefit 
immensely from the optimal join order. For distributed joins, the network cost
accounts for the majority of the total cost, and selecting the proper distributed join method
is critical for query performance. 
An increasing number of studies~\cite{cboReview19,cboSurvey15} have focused on comprehensive costs in distributed environments. 
Many cost models~\cite{costSpark18,distanceJoin23,queryOptCloud21,costSparkSqlJoin18,skewJoinSpark22,compJoinSpark21} also highlight the network cost;
however, they are not primarily designed 
for a scalable environment and rely on some inputs that can be easily derived in data-intensive scalable computing (DISC) platforms,
or they involve abundant hyperparameters that are specific to the complex scalable environment and face potential practical problems. 

We aim to design and implement a precise and practical cost model for distributed joins 
that appropriately addresses the network and computing costs of DISC platforms. 
We propose \methodName, which is an adaptive cost-based query optimization strategy for 
selecting distributed join methods when deciding on a physical query plan, 
and seamlessly integrate it as a physical plan optimization rule in SparkSQL, 
which is an industry-prevalent big data processing platform. 
The \methodName\ source code is open for access.
It has the following features that are specifically designed to tackle problems with cost-based distributed join method selection.
First, the cost model incorporates the network and computing workloads
of different joining phases 
of different physical distributed join methods in the cluster.
The cost model is precise and simple with only one hyperparameter to weight the network workload
against the computing workload, 
where the network workload depends on the data sizes, communication pattern,
and cluster configuration. 
Second, the statistics of the datasets that are used for the join method cost model
are adapted from actual runtime statistics of every data exchange phase, 
which enables close-to-fact cost estimation.
During the data exchange phase of every join or group-by-like operation, 
the query optimizer collects the runtime statistics of the output dataset
including the cardinality and size. 
The optimizer adapts the output statistics of the
successor operations based on the phase-updated  
runtime statistics and uses the adapted runtime statistics for the cost calculation
in the subsequent join operation.
Third, the physical query plan can be efficiently re-optimized during its execution
based on the adaptive runtime statistics. 
With such statistics adaptively updated
based on the runtime results, 
the physical query plan is re-optimized by re-selecting
better physical methods for the remaining joins.

The evaluation results show that \methodName\ accurately selects the distributed join methods and thus
significantly improves the performance of join queries in TPC-DS~\cite{tpcdsUrl}, 
which is an industry-standard database test suite for big data processing platforms.
Furthermore, the relative size of the joining datasets is a better criterion than 
the absolute size for judging
whether a broadcast or shuffle join method costs less and is hence a preferable option for a logical join. 
We further propose the \textit{Performance Sensitivity To Selection} (PSTS)
as a general metric to measure the optimization effectiveness of join method selection
strategies in different benchmarks, 
and show that \methodName\ effectively optimizes the query plan 
by selecting the best join methods.
The major contributions of this study are summarized as follows:
\begin{itemize}
	\setlength{\itemsep}{0pt}
	\setlength{\parskip}{0pt}
	\setlength{\parsep}{0pt}
	
	\item We build a cost model of various distributed join methods 
	with different communication patterns and in-memory operations
	and highlight the importance of the relative size of joining datasets 
	when comparing different distributed join methods. 
	Our cost model is robust in data skew scenarios and particularly practical for the complex distributed environment 
	as it has only one hyperparameter 
	which is the relative weight of the network cost of the distributed join method.
	
	\item We propose \methodName\ query optimization strategy 
	and seamlessly implement it in SparkSQL. 
	This strategy selects the best distributed join methods based on adaptive runtime statistics and efficiently re-optimizes the physical query plan. 
	
	\item Our evaluation results show that \methodName\
	can reduce the TPC-DS query completion time by 21\% 
	and achieve the best performance in 62 of the 97 queries
	compared with other available join method optimization strategies. 
	We also propose the novel PSTS
	as a general metric for measuring the optimization effectiveness
	of various join method selection strategies. 
\end{itemize}

The remainder of this paper is organized as follows. 
Section~\ref{section:background} presents the background on distributed join methods
and distributed query optimization.
In Section~\ref{section:costModel}, 
we model, analyze, and compare the costs of various distributed join methods
in different joining phases. 
We describe the optimization framework of \methodName\
and the implementation details in Section~\ref{section:framework}
and present the evaluation results for the TPC-DS benchmark in Section~\ref{section:evaluation}.
Section~\ref{section:relatedWork} outlines the related work, 
and Section~\ref{section:conclusion} concludes the paper and notes the plans for future work.

\section{Background}\label{section:background}

\subsection{Distributed join methods}
Various physical implementation methods are available for a logical join operation
depending on the computing paradigm. In contrast to local join operations
that join data within a computing node, 
the implementation of distributed operations involves an additional data exchange phase that rearranges
data locations across nodes via the network prior to the join.
Distributed join methods vary according to their implementation methods in the data
exchange and local join phases.
For simplicity, 
we use the word ``join'' to refer to a logical join operation
or physical join method, depending on the context.

\subsubsection{Broadcast and shuffle} 
Two main methods are used in the data exchange phase.
The broadcast method sends an entire replicate of the smaller table on one side of the join operation
to all nodes in the cluster, whereas the data of the other table remain unmoved. 
This can be viewed as the cross-node version of replicating small tables in multiple cores~\cite{adaptiveBig13}.
The shuffle method partitions both tables based on the hash values of the join keys
and relocates their data partitions to the corresponding nodes.
The network costs are a major concern when comparing these two methods.
In general, the broadcast method may perform better when the table to be broadcast 
is sufficiently small, whereas the shuffle method is preferred when both tables are large.
The data rows that are required for a local join at each node
should be ready following the data exchange phase. 

\subsubsection{Hash, sort, and nested loop} 
The hash, sort, and nested loop are typical methods for the local join phase that handle
the partitioned table data at each node.
The hash join~\cite{hashJoin11} first builds a hash map for the smaller table and then probes the hash map 
to identify the matching rows for each row in the larger table. 
This requires additional memory space to store the hash map.
The sort join~\cite{sortMerge12}, which also known as the sort-merge join, first sorts both tables using the join key 
and then merges the two sorted tables. 
Both the hash and sort methods are only used 
in equi-joins, and the sort method requires the join key to be sortable.
The nested loop join simply uses two loops over all rows in both tables to identify the matching rows.
This method is inefficient and should not be selected unless the hash and sort joins are infeasible.

There is a debate over which of the hash and sort methods is more efficient. 
The answer depends on various factors including the CPU and memory architecture
and the in-memory implementation. In generally, a hash join 
is considered as the default join method~\cite{hashJoin14,sortVsHash09}. 
The sort join performs better when wide SIMD instructions 
and NUMA access are available~\cite{sortVsHash13}.
Our cost model ranks 
the hash method higher than the sort method (Section~\ref{section:costComparison});
however, our results (Section~\ref{section:evalAccuracy}) show that the performance difference between
them is so small that the choice can be flexible.

In this study, we focus on discussing only certain often-used distributed join methods,
such as the broadcast hash join, shuffle hash join and shuffle sort join. 
We find that the choice between broadcast and shuffle has
a much greater impact on the performance than the choice between hash and sort. 

\subsection{Distributed query optimization}\label{section:queryOpt}
A query is parsed into
a tree structure to direct the execution of a series of data transformation
operations for logical and physical query plans in many distributed database systems~\cite{sparkSql15,hive09}.
To achieve better performance, the query optimizer analyzes, rewrites, and optimizes 
both the logical and physical plans by applying 
some well-studied optimization rules and cost models~\cite{subqueryOpt09,queryOpt15}. 
In this study, we use the terms query plan and plan interchangeably. 

A logical plan is an abstract picture of the structure and sequence of a series of logical operations,
such as join, filter and groupBy. It determines the output result of a query but does not specify
the implementation method for each operation. 
Various intermediate logical plans that can output the same query results are generated; 
however, only the optimal plan is eventually selected to generate the physical plans.

Once the optimized logical plan is ready, the system translates it into a physical plan
that defines the detailed physical methods used for each logical operation.
The plan optimizer can directly determine the best physical method for a logical operation
by following pre-defined rules. 
The optimizer may also generate a set of physical plans 
each of which represents a possible combination of physical methods
from the logical plan, and select the optimal physical plan based on a cost model.
Finally, the optimized physical plan is used to generate code that can be executed in
the distributed database system. However, it may also be re-optimized at runtime using 
updated statistics or configurations. 


In this study, we use a precise cost model to select the best distributed join methods 
for logical join operations and to optimize the physical plan. 
Using the cost model to optimize join orders is a traditional topic under
local joins, which has been studied extensively in the context of traditional
database systems, and thus, is not the focus of this study.

\subsection{Adaptive statistics and the cost model}\label{section:bgStatistics}
The cost model for the query optimizer requires the statistics of the input datasets to estimate the operational costs.
Using the statistics of the input datasets, the optimizer can statically analyze the statistics of 
the output dataset for every operation along the logical plan~\cite{costOracle06}.
These statistics usually include the cardinality, size, and some column-specific statistics, 
such as the distributions of column values. The analyzed statistics are approximate estimates
and may deviate significantly from the actual dataset following a sequence of operations along the query plan.
Biased statistics may result in incorrect optimization decisions.

Adaptive runtime statistics~\cite{adaptiveStat17} provide more close-to-fact estimates
of the cost model than statically analyzed ones.
Distributed computing paradigms~\cite{sparkSql15,hive09} typically separate the query plan into
several query stages, where the data exchange phase of a shuffle-like operation
is the synchronization barrier between
the query stages of parallel tasks in all nodes. 
The query plan optimizer can update the statistics of the datasets to actual runtime statistics 
during the execution of every data exchange phase
so that the dataset statistics are expected to be runtime-accurate at the beginning of every query stage.
The optimizer uses the updated runtime statistics as a starting point to estimate 
the statistics of the output datasets for successor operations in the new query stage,
which we refer to as adaptive runtime statistics in this study.
Our proposed cost model uses adaptive runtime statistics as the inputs, 
leading to precise cost estimates for
distributed join methods and optimal method selection decisions.  
The physical plan is re-optimized based 
on the phase-updated adaptive runtime statistics.

Join cost models~\cite{distributedJoin17, forecastCost15} usually formulate 
the costs of different types of resources (e.g., disk, memory, and CPU)
or access patterns (e.g., sequential and random access)
in different execution phases. 
The cost of the distributed join methods should consider
the network cost in the data exchange phase and other computing costs 
in the local join phase.

In this study, we use adaptive runtime statistics to formulate the costs 
of the different phases of various distributed join methods, with special consideration
of the network cost.

\section{Cost models}\label{section:costModel}
In this section, we discuss the modeling assumptions and model the costs of various distributed join methods. 
We provide insights into the cost comparisons of these distributed join methods and the data skew problem.

\subsection{Modeling assumptions}

\subsubsection{Metrics} 
Certain studies have modeled the 
query completion time~\cite{costSpark18,distributedJoin17} 
or workload of a node with multiple cores~\cite{forecastCost15}
as a metric to estimate the cost of a join method. 
In this study, we use the cluster workload of different phases, 
which measures the volumes of the input datasets and related computational and communication complexities in the cluster,
as the cost metric with the following considerations:
First, the workload estimation is simple but more accurate.
It does not require detailed information such as 
the processing speed of all related computing, I/O resources, and access patterns, 
which vary in different environments and are usually difficult to represent precisely.
Second, data skew is not a severe problem in cluster workload estimation.
By summing up the workload of each node with skewed data, the skewed cluster workload
would be close to the cluster workload as if the data were evenly distributed in the cluster, 
especially when the cluster has many parallel tasks.
As a comparison, the completion time estimation is determined by the last
finished task, which is known as the straggler, and requires the global data distribution statistics.
Third, as the paradigms of various distributed join methods can be uniformly abstracted to the 
combination of data
exchange and local join
phases, where the performance of each phase is dominated by the same resource type
(i.e., computational and communicational),
the cluster workloads of each phase of the different distributed join methods are easily comparable.

\subsubsection{Workload} 
The workload in the data exchange phase is the network communication workload, 
or network workload.
The workload in the local join phase involves the use of computing resources in the local node, 
including the CPU, memory, and disk if data are spilled or virtual pages are swapped.
We summarize this workload as the computing workload.

\subsubsection{Data distribution} 
Collecting the global distribution histogram statistics is expensive.
Fortunately, with the cluster workload as the metric, assuming that the data are uniformly distributed
would generate an approximate result that is sufficient for cost comparison~\cite{rackRdmaJoin15,distributedJoin17}, 
while simplifying the cost model and reducing the optimization overhead.
We first assume a uniform data distribution across nodes 
in the cluster in each phase
and use the cluster workload to measure the cost of the distributed join methods. 
In Section~\ref{section:skew}, we later analyze the influence of skewed data on the cost model to show that 
our model is invariant when data skew exists.

\begin{table}[!t]
	\caption{Model notations}\label{table:var}
	\begin{center}
		\begin{tabular}{clcl}
			\Xhline{2\arrayrulewidth}
			\textbf{Notation} & \textbf{Description} & \textbf{Notation} & \textbf{Description} \\
			\Xhline{2\arrayrulewidth}
			$|A|$ & the size of $A$ &  $|B|$ & the size of $B$\\
			$a$ & the cardinality of $A$ &  $b$ & the cardinality of $B$\\
			$|A| / a$ & the row size of $A$ &  $|B| / b$ & the row size of $B$\\
			$|A| / p$ & the partition size of $A$ &  $|B| / p$ & the partition size of $B$\\
			$a / p$ & \multicolumn{3}{l}{the cardinality of a partition of $A$} \\
			$b / p$ & \multicolumn{3}{l}{the cardinality of a partition of $B$} \\
			$l_{fan}$ & \multicolumn{3}{l}{the average \# of matching rows in $B$ for each row in $A$}\\
			
			$w$ & \multicolumn{3}{l}{the relative weight of the network cost against the} \\
			 & \multicolumn{3}{l}{computing cost} \\
			$k$ & \multicolumn{3}{l}{the relative size coefficient of $A$ and $B$, where $|A| = k|B|$} \\
			$n$ & \multicolumn{3}{l}{the number of cluster nodes}\\
			$p$ & \multicolumn{3}{l}{distributed join parallelism}\\
			\Xhline{2\arrayrulewidth}
			\textbf{Notation} & \multicolumn{3}{l}{\textbf{The cluster (network/computing) workload of ...}}\\
			\Xhline{2\arrayrulewidth}
			$C_{broadcast}$ & \multicolumn{3}{l}{(network) broadcasting a dataset in broadcast hash join}\\
			$C_{build}$ & \multicolumn{3}{l}{(computing) building the hash map in broadcast hash join}\\
			$C_{probe}$ & \multicolumn{3}{l}{(computing) probing the hash map in hash joins}\\
			$C_{shuffle}$ & \multicolumn{3}{l}{(network) shuffling partitions in shuffle joins}\\
			$C_{sort}$ & \multicolumn{3}{l}{(computing) sorting partitioned data in shuffle sort join}\\
			$C_{merge}$ & \multicolumn{3}{l}{(computing) merging sorted partitions in shuffle sort join}\\
			$C'_{build}$ & \multicolumn{3}{l}{(computing) building the hash map in shuffle hash join} \\
			$C_{NL}$ & \multicolumn{3}{l}{(computing) nested loop in broadcast nested loop join} \\
			\Xhline{2\arrayrulewidth}
		\end{tabular}
	\end{center}
\end{table}

\subsubsection{Problem setting}
We model the cost of a distributed equi-join of datasets $A$ and $B$, 
which runs in a cluster of $n$ nodes, where each node runs several 
executors to perform the data exchange and local join tasks in parallel. 
Each task handles a partition of the dataset.
Suppose that the number of shuffle partitions for the join is $p$, 
which is known as the distributed join parallelism. 
The notations are listed in Table~\ref{table:var} for ease of reference.
Assume that
$|A| \ge |B|$. 
We use the larger dataset to join the smaller dataset. 
Thus, datasets $A$ and $B$ are located on the left and right sides of the join, respectively.
The cardinalities of $A$ and $B$ are substantially higher than the distributed join
parallelism; that is, $a \gg p$ and $b \gg p$.
Both datasets are uniformly distributed to the partitions. 
The average number of matching rows in dataset $B$ for each row in dataset $A$, 
which is known as the fanout level of $A$, 
is denoted as $l_{fan}$.
We assume that the join keys are sortable so that the sort join is feasible.

\begin{figure}[!t]
	\centering
	\includegraphics[width = 0.8 \columnwidth]{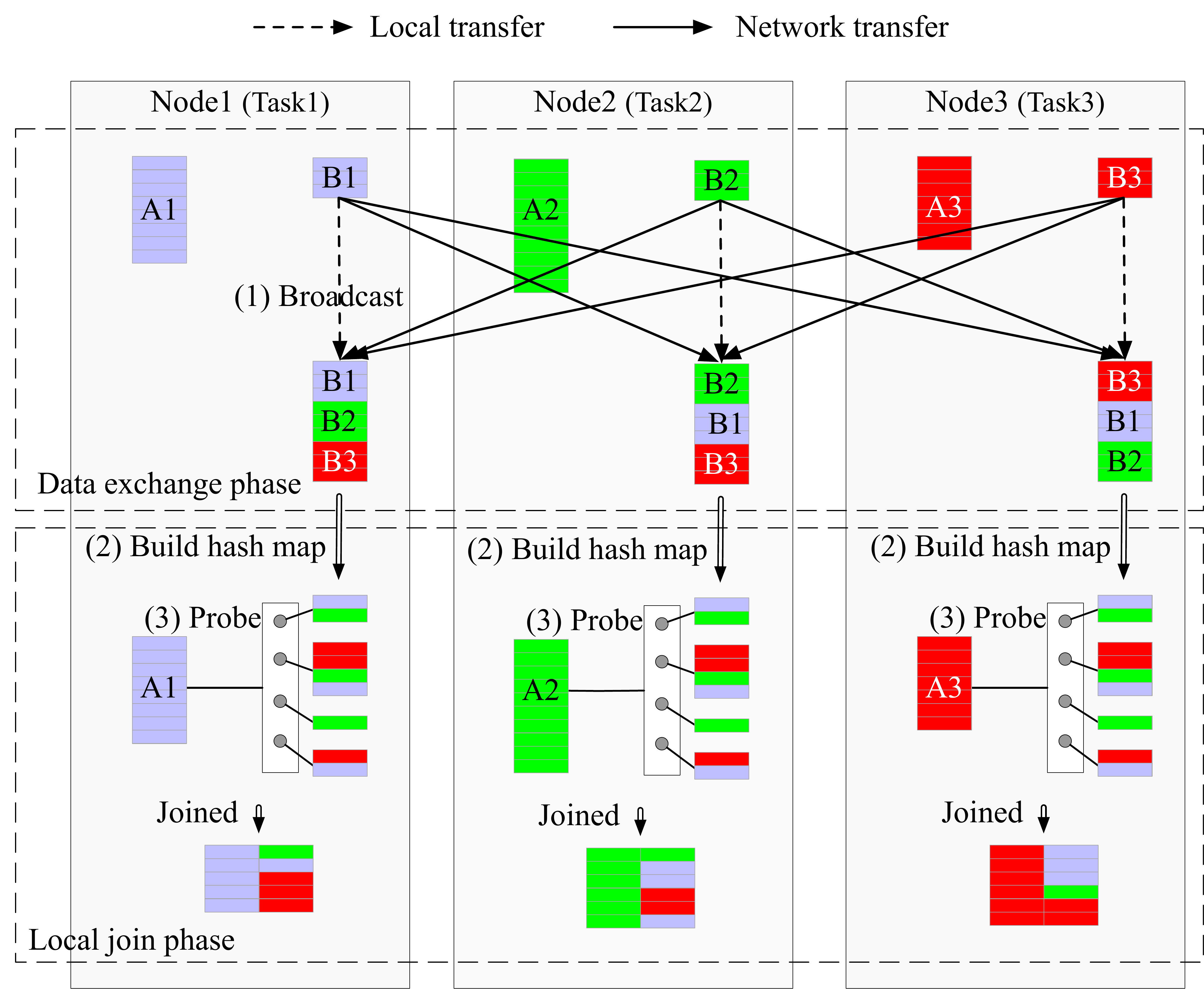}
	\caption{The broadcast hash join}
	\label{fig:broadcastHash}
\end{figure}

\subsection{Broadcast hash join}
Fig.~\ref{fig:broadcastHash} depicts the broadcast hash join procedure for datasets $A$ and $B$, with three
nodes in the cluster and one task in each node. 
It comprises three steps: broadcast, 
build, and probe.

\subsubsection{Broadcast}
In the data exchange phase, all $p$ tasks collect the entire replicate of the smaller dataset $B$.
In each task, $1/p$ of the dataset already resides in that task process and can be 
used directly (the dashed arrows of Step 1 in Fig.~\ref{fig:broadcastHash}), 
and $(p - 1)/p$ of the dataset is retrieved from other tasks 
either in the same node or nodes across the network (the solid arrows of Step 1 in Fig.~\ref{fig:broadcastHash}). 
Although fetching a partial dataset from other tasks in the same node 
via the network interface can be optimized via the local disk or shared memory access, 
this is usually considered network I/O in practice.
Therefore, the workload of each task in the broadcast step is $|B|(p - 1)/p$, 
and the cluster network workload of $p$ tasks is
\begin{equation} \label{eq:broadcastCost}
	C_{broadcast} = (p - 1)  |B|.
\end{equation}

\subsubsection{Build}
In the local join phase, 
each task first builds a hash map 
for the entire dataset $B$ (Step 2 in Fig.~\ref{fig:broadcastHash})
to accelerate the later probe step. 
The workload for building the hash map for each task is $|B|$,
and the cluster computing workload is
\begin{equation} \label{eq:buildCost}
	C_{build} = p |B|.
\end{equation}

\subsubsection{Probe}
In the probe step, the task traverses each row in a partition of dataset $A$,
where the partition cardinality is $a / p$,
and probes the hash map to identify matching rows 
with the same key value in the entire dataset $B$ (Step 3 in
Fig.~\ref{fig:broadcastHash}).
The workload of each task depends on 
the fanout level of $A$, $l_{fan}$.
It visits rows of size $|A| / a$  $a / p$ times 
and those of size $|B|/b$ $a l_{fan}/p$ times.
The cluster workload of the probe is 
\begin{equation*}
	C_{probe} = p  (\frac{a}{p} \frac{|A|}{a} + \frac{al_{fan}}{p} \frac{|B|}{b})
	= |A| +  \frac{a l_{fan}}{b}|B|.
\end{equation*}
In the best case, when no rows in dataset $A$ can find 
a matching row and $l_{fan} = 0$, the cluster workload is simply traversal of all rows in dataset $A$
and is $|A|$.
In the worst case, when the key values of all rows in datasets $A$ and $B$ are
the same and each row on the left of the join matches all rows on the right, $l_{fan} = b$.
The cluster workload is $|A| + a|B|$. 
On average, we can assume that the matching rows of all $a$ rows in $A$ are evenly 
distributed across all $b$ rows in $B$, 
and we obtain $l_{fan} = b / a$. Therefore,  
\begin{equation} \label{eq:probeCost}
	C_{probe} = |A| + |B|
\end{equation}

\subsubsection{Overall}
Recall that the broadcast workload is the network cost, and the build and probe workloads are the
computing costs. We assign different weights to these costs of different resource types
when combining them to obtain the overall cost.
As only two resource types are considered, instead of assigning each cost with 
a weight coefficient, that is, $w_1$ and $w_2$, 
we use a single variable $w$ to denote the relative weight of the network cost against the computing cost
for simplicity. 
Using Equations~\ref{eq:broadcastCost}, \ref{eq:buildCost}, and \ref{eq:probeCost},
the overall cost of the broadcast hash join is 
\begin{equation} \label{eq:broadcastHashCost}
	\begin{aligned}
		C_{broadcastHash} &= wC_{broadcast} + C_{build} + C_{probe}\\
		&= |A| + (wp - w + p + 1) |B|.
	\end{aligned}
\end{equation}
This is a linear function of the sizes of datasets $A$ and $B$.

\begin{figure}[!t]
	\centering
	\includegraphics[width = 0.8 \columnwidth]{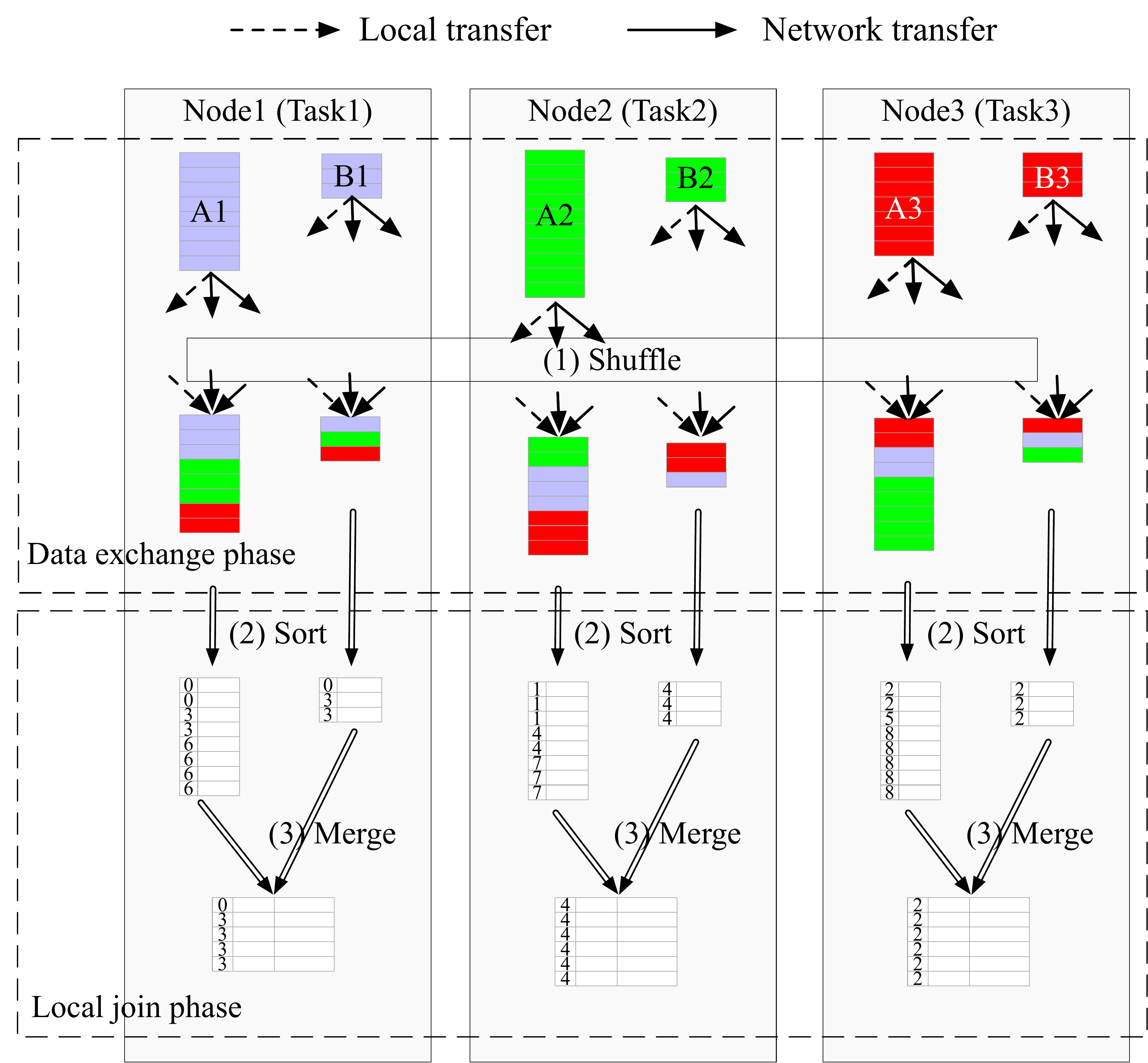}
	\caption{The shuffle sort join}
	\label{fig:shuffleSort}
\end{figure}

\subsection{Shuffle sort join}
Fig.~\ref{fig:shuffleSort} depicts the procedure of the shuffle sort join  
of datasets A and B with three nodes in the cluster and one task in each node.
This process comprises three steps: shuffle, sort, and merge. 

\subsubsection{Shuffle}\label{section:shuffle}
In the shuffle step, both datasets are repartitioned across tasks 
via the network (Step 1 in Fig.~\ref{fig:shuffleSort}).
As in the broadcasting, $1/p$ of the partition data 
are process-native and $(p-1)/p$ are retrieved via the network in each task.
The difference is that in the broadcast step, each task needs to collect the entire dataset $B$,
whereas in the shuffle step, each task only collects a partition of datasets $A$ and $B$.
Therefore, the workload of each task in the shuffle step is 
$((p-1) / p) (|A|/p + |B|/p)$. 
The cluster network workload of $p$ tasks is 
\begin{equation} \label{eq:shuffleCost}
	C_{shuffle} = \frac{p-1}{p} (|A| + |B|).
\end{equation}

\subsubsection{Sort}
In the sort step, each task sorts a partition of datasets $A$ and $B$ (Step 2 in Fig.~\ref{fig:shuffleSort}), respectively.
To sort a partition of $A$ with $a / p$ rows where the size of each row is $|A| / a$, 
the workload is $\frac{|A|}{a} \frac{a}{p} \log_2 \frac{a}{p}$, 
or $\frac{|A|}{p} \log_2 \frac{a}{p}$. 
The workload for sorting a partition of $B$ can be derived in a similar manner, 
and the workload of each task in the sort step is 
$\frac{|A|}{p} \log_2 \frac{a}{p} + \frac{|B|}{p} \log_2 \frac{b}{p}$.
The cluster computing workload is
\begin{equation} \label{eq:sortCost}
	C_{sort} = |A| \log_2 \frac{a}{p} + |B| \log_2 \frac{b}{p}.
\end{equation}

\subsubsection{Merge}
To merge two sorted partitions of datasets $A$ and $B$, 
the workload is simply the sum of the sizes of these two partitions; 
that is, $|A| / p + |B| / p$ for each task. 
The cluster computing workload in the merge step is
\begin{equation} \label{eq:mergeCost}
	C_{merge} = |A| + |B|.
\end{equation}

\subsubsection{Overall}
Similar to the overall cost of the broadcast hash join,
we assign a weight $w$ to the network cost in the shuffle phase.
Using Equations~\ref{eq:shuffleCost}, \ref{eq:sortCost}, and \ref{eq:mergeCost}, 
the overall cost of the shuffle sort join is
\begin{equation} \label{eq:shuffleSortCost}
	\begin{aligned}
		&C_{shuffleSort} = wC_{shuffle} + C_{sort} + C_{merge} \\
		&= (\frac{wp-w+p}{p} +  \log_2 \frac{a}{p}) |A| + (\frac{wp-w+p}{p} +  \log_2 \frac{b}{p}) |B|.
	\end{aligned}
\end{equation}
The result depends not only on the sizes of both datasets, 
but also on the cardinalities owing to sorting. 

\subsection{Shuffle hash join}
For the shuffle hash join, the shuffle step is identical to that 
of the shuffle sort join, as is the corresponding cluster workload, 
which is $C_{shuffle}$ in Equation~\ref{eq:shuffleCost}. 
The local join phase follows the same 
pattern as the broadcast hash join; 
however, the input sizes of the datasets for the build and probe steps differ.

In the build step for each task, instead of building the hash map
for the entire dataset $B$ of size $|B|$ as in the broadcast hash join, 
the task only needs to build the hash map for a partition of dataset $B$
which is of size $|B| / p$. Therefore, the cluster computing workload
of this build step is
\begin{equation} \label{eq:partitionBuildCost}
	C'_{build} = |B|.
\end{equation}

In the probe step, although the size of the hash map in a task is only 
$1 / p$ of that in the broadcast hash join, 
the workload depends only on the cardinality of the partition of $A$, 
row sizes of $A$ and $B$, and fanout level of $A$. 
Therefore, the cluster probe workload of the shuffle hash join is the same as that
of the broadcast hash join, which is $C_{probe}$ in Equation~\ref{eq:probeCost}. 

Using Equations~\ref{eq:probeCost}, ~\ref{eq:shuffleCost}, and \ref{eq:partitionBuildCost},
with weight $w$ assigned to the shuffle workload, the overall cost of the 
shuffle hash join is
\begin{equation} \label{eq:shuffleHashCost}
	\begin{aligned}
		C_{shuffleHash} &= wC_{shuffle} + C'_{build} + C_{probe} \\
		&= \frac{wp-w+p}{p} |A| +\frac{wp-w+2p}{p} |B|.
	\end{aligned}
\end{equation}
It is also a linear function of the sizes of datasets $A$ and $B$. 

\subsection{Broadcast nested loop and Cartesian product joins}
We introduce two variants of distributed join methods
that are used only when the above methods are infeasible, 
and discuss the reason for their poor performance based on the cost model.
These are the broadcast nested loop join (broadcast NL join)
and Cartesian product join.
They should not be selected except when the join is not an equi-join,
the join key is not sortable, or there is insufficient space to build the hash map. 

For the broadcast NL join, the data exchange phase is the same 
as that of the broadcast hash join, which is the broadcast step.
In the local join step, each task loops over all rows of 
a partition of dataset $A$, and for this every row, 
loops over all rows in the entire replicate of dataset $B$.
The outer loop runs $a / p$ times and visits rows of size $|A| / a$. 
The inner loop runs $ab/p$ times and visits rows of size $|B| / b$.
The cluster NL workload is
\begin{equation} \label{eq:nlCost}
	C_{NL} = p(\frac{a}{p} \frac{|A|}{a} + \frac{ab}{p} \frac{|B|}{b}) 
	= |A| + a|B|.
\end{equation}

We can easily determine why the broadcast NL join method is 
inferior to other distributed join methods by comparing their local join workloads.
For example, according to Equations~\ref{eq:buildCost} and \ref{eq:probeCost},
the cluster workload of the local join phase of the broadcast hash join is
$C_{build} + C_{probe}$; that is, $|A| + (p + 1) |B|$. 
As the dataset cardinality $a$ is assumed to be significantly larger than $p$, 
$C_{NL} \gg C_{build} + C_{probe}$.

Using Equations~\ref{eq:broadcastCost} and \ref{eq:nlCost}, the overall cost of the broadcast NL join is
\begin{equation*} \label{eq:broadcastNLCost}
	\begin{aligned}
		C_{broadcastNL} &= wC_{broadcast} + C_{NL} = |A| + (wp - w + a)|B|.
	\end{aligned}
\end{equation*}

Logically, the Cartesian product join is a shuffle NL join. 
The data exchange phase is a shuffle step and its cluster workload
is $C_{shuffle}$.
The NL step is similar to that of the broadcast NL join; however, the inner loop
only visits a partition of dataset $B$, which has $b / p$ rows instead of ${b}$ rows. 
Therefore, the inner loop runs  $a/p \times b/p$ times instead of $ab/p$ times.
Everything else remains the same as in the broadcast join in the NL step, 
and the cluster NL workload of the Cartesian product join is 
\begin{equation} \label{eq:shuffleNLCost}
	C'_{NL} = p(\frac{a}{p} \frac{|A|}{a} + \frac{ab}{p^2} \frac{|B|}{b}) 
	= |A| + \frac{a}{p}|B|.
\end{equation}

We can also find that the Cartesian product join is inferior 
to the shuffle hash join by comparing their local join workloads. 
According to Equations~\ref{eq:probeCost} and \ref{eq:partitionBuildCost}, 
the local join workload of the shuffle hash join is $C'_{build} + C_{probe}$, 
which is $|A| + 2|B|$. When $a/p \gg 2$, $C'_{NL} \gg C'_{build} + C_{probe}$.

According to Equations~\ref{eq:shuffleCost} and \ref{eq:shuffleNLCost}, 
the overall cost of the Cartesian product join is
\begin{equation*} \label{eq:cartesianCost}
	\begin{aligned}
		C_{cartesian} &= wC_{shuffle} + C'_{NL} \\
		&= \frac{wp - w + p}{p} |A| + \frac{wp - w + a}{p} |B|.
	\end{aligned}
\end{equation*}

\subsection{Comparison}\label{section:costComparison}
We discuss and provide insights into the cost models of the broadcast hash join, 
shuffle hash join, and shuffle sort join.

\subsubsection{Shuffle hash vs. shuffle sort}
Although the hash and sort method are considered as comparable join methods, 
in our cost model, the hash method always has a lower workload than the sort method. 
It can be inferred from Equations~\ref{eq:probeCost}, \ref{eq:sortCost}, \ref{eq:mergeCost}, and \ref{eq:partitionBuildCost}
that $C'_{build} + C_{probe} < C_{sort} + C_{merge}$.
Therefore, provided that there is sufficient memory to create a hash map for a partition of $B$, 
our cost model always prefers the shuffle hash join.
However, note that the additional memory space that is occupied by the shuffle hash join is not considered in the cost model.

\subsubsection{Broadcast hash vs. shuffle hash}
The workloads of these two methods differ not only in the data exchange phase
but also in the local join phase. These can be compared by simply calculating their overall
costs. As both costs are linear functions of $|A|$ and $|B|$, we can reformulate 
the costs by exploring the relative sizes of datasets $A$ and $B$. 

Let $|A| = k |B|$, where $k$ represents the relative size coefficient of datasets $A$ and $B$. 
If $k = k_0$ and $C_{broadcastHash} = C_{shuffleHash}$, 
solving this equation 
using Equations~\ref{eq:broadcastHashCost} and \ref{eq:shuffleHashCost}, 
we have 
\begin{equation} \label{eq:k}
	k_0 = \frac{pw + p - w}{w}.
\end{equation}
When $k \le k_0$, $C_{broadcastHash} \ge C_{shuffleHash}$.
When $k > k_0$, $C_{broadcastHash} < C_{shuffleHash}$.
The value of $k_0$ is proportional to the value of $p$.
The relative costs of these two join methods are directly determined by 
the relative size coefficient $k$, distributed join 
parallelism $p$, and network cost relative weight $w$.

This result is consistent with our intuition regarding the costs of the two distributed join methods.
When $A$ is not sufficiently large relative to $B$, 
the workload of shuffling both datasets is not heavy compared 
with that of broadcasting $B$. Furthermore, 
the shuffle method builds a smaller hash map than the broadcast method does. 
In this case, the shuffle hash join may is preferred.
However, when $A$ is sufficiently larger than $B$, with a relative size coefficient
that is larger than a particular threshold, shuffling the large dataset $A$ can become
a heavier workload than broadcasting the entire replicate of $B$
and building the hash map for it.
In this case, the broadcast hash join is preferred.
$k_0$ is the threshold at which $A$ is considered sufficiently bigger than $B$. 
When the number of parallel join tasks is larger, the cost of broadcast $B$
is greater, and the threshold should be higher for $A$ to be considered large. 
As mentioned in Section~\ref{section:queryOpt}, SparkSQL
uses a manually set absolute size to determine whether a dataset is sufficiently small
for broadcasting. Our cost model indicates that the relative size is
the criterion for selecting the broadcast or shuffle method.


\subsection{Analysis of data skew}\label{section:skew}
The cost model assumes a uniform data distribution across the nodes. 
Although considering data skew complicates the cost model, but we can briefly
discuss its influence on the current cost model if data skew is considered. 
Inferior nested-loop-like joins are not discussed here. 

The overall cluster workload of the broadcast hash join ($C_{broadcastHash}$ (\ref{eq:broadcastHashCost})) remains unchanged.
When the cluster workload of a step of a distributed join explains the expected sum of the partition data sizes,
with every partition weighted by the same constant value, 
the cluster workload remains the same even if the dataset is skewed across the nodes. 
This is the case with $C_{broadcast}$ (\ref{eq:broadcastCost}), $C_{build}$ (\ref{eq:buildCost}), $C_{probe}$ (\ref{eq:probeCost}), $C_{merge}$ (\ref{eq:mergeCost}), and $C'_{build}$ (\ref{eq:partitionBuildCost}).

The overall cluster workload of the shuffle hash join ($C_{shuffleHash}$ (\ref{eq:shuffleHashCost})) also remains unchanged. 
The shuffle workload is not affected by the data distribution but by the data key dependency 
between the join input and output~\cite{confluence17}. 
In Section~\ref{section:shuffle}, our cost model implicitly implies the data key dependency 
by assuming that $(p - 1)/p$ of each partition data
will be transferred via the network. 
In extreme cases, when all data rows are already in the nodes where they will be processed later, no data
must be shuffled, and $C_{shuffle} = 0$. In contrast, if all data in both datasets are not
in the nodes where they will be processed and the entire dataset needs to be transferred via the network,
$C_{shuffle} = |A| + |B|$. 
In the data skew case, if the same data key dependency assumption is held, 
the cluster workload of the shuffle step $C_{shuffle}$ (\ref{eq:shuffleCost}) is the same as that of the uniformly distributed case.

The sort-step workload and thus the overall cluster workload of the shuffle sort join will be higher. 
The cluster workload of the sort step contains the expected sum of the products of
the partition cardinality and the logarithm of the partition cardinality,
which has the minimum value when the cardinalities of all partitions are the same in the uniform distribution case.
When the data are skewed, it is higher than $C_{sort}$ (\ref{eq:sortCost}). 
The higher cost of the shuffle sort join in the data skew case does not change its rank 
compared to the shuffle hash join, which is preferred by our model. 

That is, considering data skew does not affect the comparable costs
of the distributed join methods in our cost model.

\begin{table}[!t]
	\caption{A summary of the costs of different distributed join methods with the suffixes of $C$ indicating the method types and their ranks (higher-rank methods are preferred when they are feasible)}\label{table:costSummary}
	\begin{center}
		\begin{tabular}{c l c}
			\Xhline{2\arrayrulewidth}
			\textbf{Join method} & \textbf{Cost} & \textbf{Rank}\\
			\Xhline{2\arrayrulewidth}
			$C_{broadcastHash}$ & $ |A| + (wp - w + p + 1) |B|$ & 3\\
			$C_{shuffleSort}$ & $(\frac{wp-w+p}{p} +  \log_2 \frac{a}{p}) |A| + (\frac{wp-w+p}{p} +  \log_2 \frac{b}{p}) |B|$ & 2\\
			
			$C_{shuffleHash}$ & $\frac{wp-w+p}{p} |A| +\frac{wp-w+2p}{p} |B|$ & 3\\
			
			$C_{broadcastNL}$ & $|A| + (wp - w + a)|B|$ & 1\\
			$C_{cartesian}$ & $\frac{wp - w + p}{p} |A| + \frac{wp - w + a}{p} |B|$ & 1\\
			\Xhline{2\arrayrulewidth}
		\end{tabular}
	\end{center}
\end{table}

\subsection{Cost summary}
The costs and the rankings of the different distributed join methods mentioned above
are summarized in Table~\ref{table:costSummary}. Of all feasible join methods for a specific logical join, 
the join methods are selected based on the rank first and cost size later. 
The join method with the highest rank and lowest cost is selected, as explained
in the following section.

\section{Cost-based optimization framework}\label{section:framework}
In this section, we describe the query optimization framework of \methodName\
which incorporates the cost model to select the best distributed join method 
based on adaptive runtime statistics. 
We also introduce the details of the implementation of \methodName\ in SparkSQL.

\subsection{Optimization procedure}
\begin{figure}[!t]
	\centering
	\includegraphics[width = 1 \columnwidth]{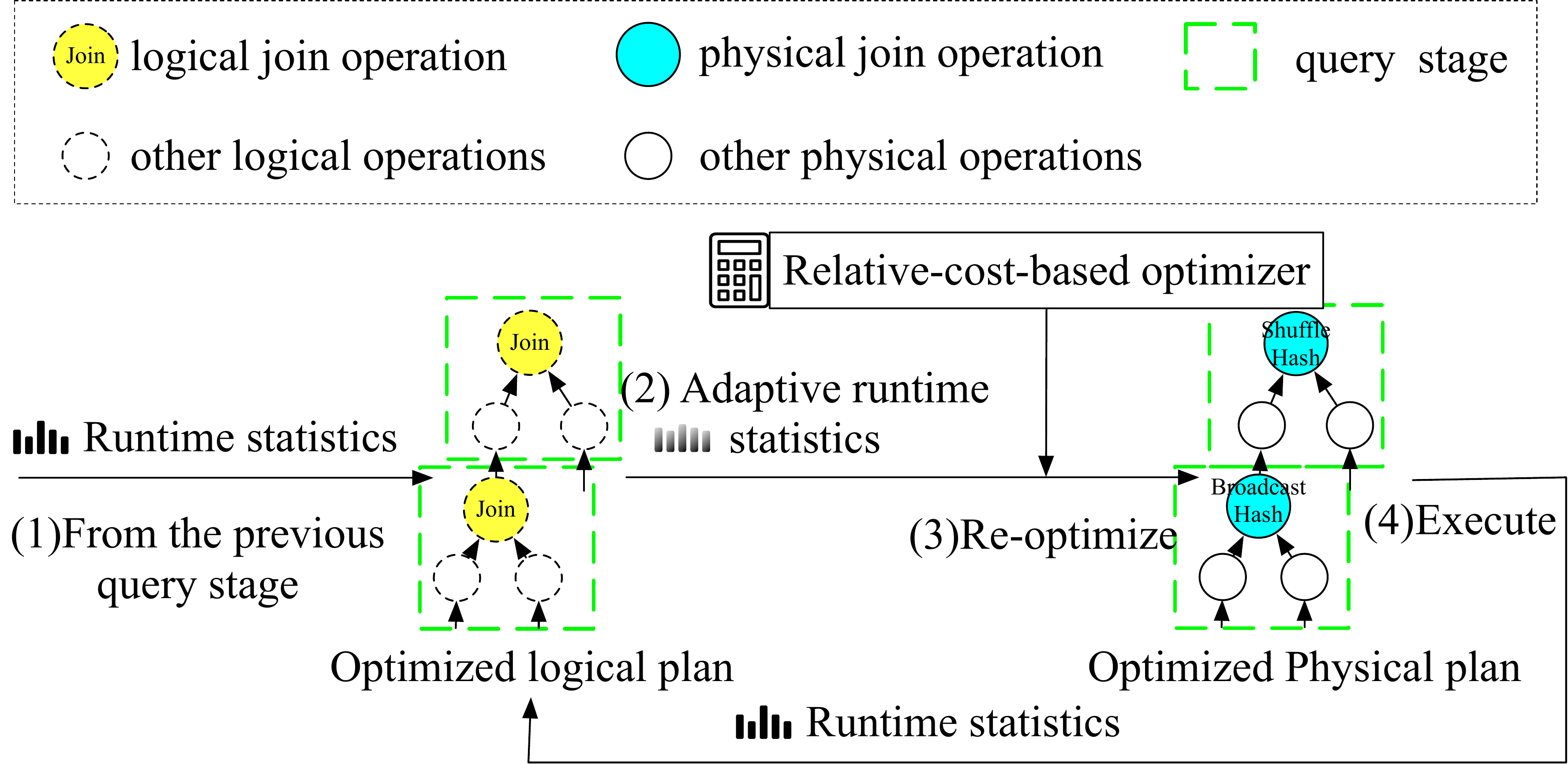}
	\caption{The adaptive cost-based join method optimization procedure}
	\label{fig:optFramework}
\end{figure}

Fig.~\ref{fig:optFramework} depicts the optimization procedure of \methodName.
The tree with dashed nodes represents an optimized logical query plan, and the tree with
solid nodes represents the (re-)optimized physical query plan.
As mentioned in Section~\ref{section:bgStatistics}, a query plan is divided into query stages. 
The nodes in the dark background are data exchange operations that act as the
boundaries of the query stages. 
In Step 1, every query stage is executed
according to the optimized physical plan,
and the runtime statistics of the corresponding operation nodes in the logical query plan are updated. 
According to our cost model, the required statistics are the size and cardinality of the output dataset.
In Step 2, the logical plan adapts the updated runtime statistics along the query path
until it reaches the next data-exchange operation, such as a join operation. 
In Step 3, for this logical join operation, the cost model uses the adaptive runtime statistics 
to select the best distributed join method for the current stage 
and generates a new optimized physical plan.
This step is known as the re-optimization of the physical plan, which we will
discuss further.
In Step 4, the execution of the distributed join collects 
the runtime statistics of the output for the re-optimization of the next stage starting from Step 2.
This procedure is iterated until it reaches the root of the query plan and retrieves the final result.

\subsection{Optimization time complexity}
The physical plan optimization for the distributed join method selection 
in \methodName\ is time efficient.
In certain cost-based optimization strategies~\cite{costOracle06}, the optimizer
generates many alternative plans, 
each of which is a possible combination of the choice results 
for a sequence of join operations.
Subsequently, it uses a cost model to select the best one with the lowest cost. 
Determining the globally optimal plan in this manner is an NP-hard problem. 
For example, for a query plan with $h$ successive joins, 
where each join has $l$ physical methods for selection, 
the time complexity for determining the globally optimal plan for the above strategy 
is $O(l^h)$. 
In our setting, where the data are uniformly distributed across the cluster, 
because the cost of a distributed join method for every logical join is independent of 
the costs of other logical joins, 
the optimization of a logical join is independent of the optimization of the others. 
In every physical plan re-optimization iteration in \methodName, 
the optimizer only independently determines the optimal 
join method for the corresponding logical join to be executed.
A globally optimal plan is generated and executed 
after $h$ optimizations or re-optimizations
of the physical plan. The time complexity 
is $O(lh)$, which is a time-efficient polynomial.

\subsection{Distributed join method selection}~\label{section:methodSelection}
We describe the algorithm for using the cost model to select 
the best distributed join method for a logical join in every iteration 
of the physical plan optimization or re-optimization.
Algorithm~\ref{algo:joinSelection} presents the brief selection 
algorithm for an equi-join, in which two datasets are joined 
based on equality or matching column values.
When no user-defined hint is provided for the selection preference, 
the optimizer calculates the costs of various distributed join methods.
It selects the join method with the lowest cost 
permitted by the join type and memory requirements.
The best distributed join method for this logical join is returned. 
In the algorithm, the shuffle hash join is ranked higher than the shuffle
sort join, and the hash- and sort-based joins are ranked higher than the NL-based joins.

If a non-equi-join is inner-like, as only the Cartesian product join and broadcast NL join are feasible, 
the optimizer selects the join method with the lower cost.

\begin{figure}[!t]
	
	\begin{algorithm}[H]
		\DontPrintSemicolon
		\KwIn{the logical equi-join}
		\KwIn{the cost model that return the cost of different distributed join methods for a logical join}
		\KwOut{the distributed join method}
		
		\If{the user-defined join hint exists in the logical join} {
			\KwRet{the hinted join method}
		}
		
		$C_{broadcastHash}$ = the broadcast hash join cost\\
		$C_{shuffleHash}$ = the shuffle hash join cost\\
		
		\uIf{$C_{broadcastHash} < C_{shuffleHash}$ \textbf{and} hashing is allowed} {
			\KwRet{the broadcast hash join}
		} \uElseIf{hashing is allowed} {
			\KwRet{the shuffle hash join}
		} \uElseIf{the join key is sortable} {
			\KwRet{the shuffle sort join}
		} \uElse {
			$C_{broadcastNL}$ = the broadcast NL join cost\\
			$C_{cartesian}$ = the Cartesian product join cost\\
			\uIf{$C_{cartesian} \le C_{broadcastNL}$ \textbf{and} the join type is inner-like} {
				\KwRet{the Cartesian product join}
			} \lElse {
				\KwRet{the broadcast NL join}
			}
		}
		
		\caption{Cost-based distributed join method selection}
		\label{algo:joinSelection}
	\end{algorithm}
\end{figure}

\subsection{Implementation details}
We implement \methodName\ in SparkSQL as a physical plan optimization rule.
Consequently, \methodName\ automatically applies the optimization mechanisms that are already implemented by SparkSQL, 
including the adaptive runtime statistics and the physical plan re-optimization.
The following implementation factor are considered: 

\begin{itemize}
	\setlength{\itemsep}{0pt}
	\setlength{\parskip}{0pt}
	\setlength{\parsep}{0pt}
	\item \textbf{Backward compatibility}: \methodName\ should be compatible with the original 
	join method selection strategy and other query plan optimization techniques. 
	\item \textbf{Data reliability}: \methodName\ should measure the join operation costs 
	and make selections based on reliable input statistics.
	\item \textbf{User friendly}: \methodName\ should be seamlessly implemented on the platform 
	and can be used by users with minimum learning effort. 
\end{itemize}

The respective implementation details for dealing with these considerations are as follows.

First, \methodName\ is implemented in the same physical plan optimization rule 
where the original distributed join method selection strategy is implemented 
such that either \methodName\ or the original strategy can run alongside other query optimization techniques.
The original join selection strategy follows an absolute-size approach, 
which prefers the broadcast method over the shuffle method 
when the size of the smaller dataset is within a user-defined absolute value, such as 10 MB.
When \methodName\ is configured to be enabled, the original strategy is switched to \methodName\ 
unless the statistics are invalid, as described below.

Second, the statistics are not always ready or meaningful for an operation, particularly 
in the first query stage. 
For example, because of the lazy execution mechanism, 
when loading a dataset from a data source that does not have statistics in the header, 
the optimizer cannot provide a proper estimate of the statistics and initializes
the dataset size statistics to a very large number. 
Using the adaptive statistics from this initialized value for the cost model
will lead to incorrect join method selection decisions, such as broadcasting and building
a hash map for a very large dataset. 
To address this problem, \methodName\ treats only size statistics below a watermark number
as valid statistics.
This watermark is the largest dataset size that is supported the \methodName\ strategy, 
and we set it to 100 GB by default. 
In the case of invalid statistics for a logical join, 
\methodName\ switches to the original join selection strategy for that join. 
The original strategy may simply think that the datasets are too large for broadcasting and hashing, 
and select the sort merge join.

Third, the weight of the network cost relative to the computing cost $w$ is user configurable.
This is the only variable in \methodName\ that depends on the environment. 
In Section~\ref{section:evalWImpact}, we demonstrate that \methodName\ performs well with different $w$ values
and users do not need to spend much time tuning this parameter. 

\section{Evaluation}\label{section:evaluation}
In this section, we present the performance evaluation results of \methodName\ 
based on the TPC-DS benchmark and compare them with those of other join method selection
strategies.
The results show that \methodName\ improves the query time significantly by selecting 
the best distributed join methods. We statistically analyze the query plans 
and discuss the insights and key factors that make \methodName\ a better 
optimization strategy. 
We also explore the impact of different values of the relative weight of the network cost on the 
computing cost, $w$.

\begin{table}[!t]
	\caption{(a) Evaluation benchmark specifications (the scale, the raw text size of the whole benchmark,
	and the raw text size of the largest dataset in the benchmark),  
	testbed executor and node resource capacity configurations; 
	(b) excluded queries for the benchmarks of each scale; 
	and (c) strategies for distributed join method selection}\label{table:strategies}
	\begin{center}
		\begin{tabular}{c cccc}
			\Xhline{2\arrayrulewidth}
			\textbf{Scale} & \textbf{Whole size} & \textbf{Largest size} & \textbf{Node} & \textbf{Executor}\\
			1 & 1.2 GB & 386 MB & 16 GB & 4 GB/1 core\\
			10 & 12 GB & 3.7 GB & 32 GB & 14 GB/2 cores\\
			100 & 96 GB & 38 GB & 32 GB & 14 GB/2 cores\\
			\Xhline{2\arrayrulewidth}
			
			\textbf{Scale} & \multicolumn{3}{l}{\textbf{Excluded queries}}& \textbf{\# Queries}\\
			1 & \multicolumn{3}{l}{\textit{q10, q35, q69} for AQE and \methodName} & 97\\
			10 & \multicolumn{3}{l}{\textit{q10, q35, q69, q72} for all} & 93\\
			100 & \multicolumn{3}{l}{\textit{q10, q14a, q14b, q35, q37, q69, q72} for all} & 90\\
			\Xhline{2\arrayrulewidth}
			\textbf{Strategy} & \multicolumn{4}{l}{\textbf{Description}}\\
			ShuffleSort & \multicolumn{4}{l}{Force to select the shuffle sort join if keys are sortable.}\\
			
			ShuffleHash & \multicolumn{4}{l}{Force to select the shuffle hash join if the dataset is} \\ 
			            & \multicolumn{4}{l}{small enough for building the hash.}\\
			
			AQE & \multicolumn{4}{l}{Select the broadcast hash join if the size of the adaptive}\\
			&\multicolumn{4}{l}{runtime statistics of a dataset does not exceed 10 MB.}\\
			&\multicolumn{4}{l}{Otherwise, select the shuffle hash or shuffle sort join.}\\
			
			\methodName & \multicolumn{4}{l}{Network workload weight $w = 1$ and hence $k_0 = 39.$}\\
			\Xhline{2\arrayrulewidth}
		\end{tabular}
	\end{center}
\end{table}

\subsection{Benchmark, testbed, and metrics}
\textbf{Benchmark}: We use the TPC-DS benchmark~\cite{tpcdsUrl} to evaluate the performance of \methodName. 
TPC-DS is a representative benchmark with complex decision workloads 
for general big data processing systems. 
We perform testing on the TPC-DS benchmark of different size scales: 1, 10, and 100. 
The text sizes of the each benchmark and the largest dataset are listed 
in Table~\ref{table:strategies}.
We run the TPC-DS queries that are natively integrated by SparkSQL, 
which excludes some flaky queries and has a total of 97 remaining queries. 
Some queries fail to run because of the use up of the allocated memory in certain strategies;
therefore, they are excluded from the comparison of the query time results.
The excluded queries in the benchmarks of different scales are listed in Table~\ref{table:strategies}.
The datasets are transformed to the parquet format.
Each query runs three times for all tests and the average results are presented.

\textbf{Testbed}: We deploy SparkSQL running on YARN~\cite{yarn} in a cluster of six computer nodes, 
where each node is equipped with 12 CPU cores at 2.6 GHz and 64 GB of memory.
The nodes are mounted on a single rack connected by GbE ports. 
One node is configured as the HDFS name node and the YARN resource manager. 
The remaining five nodes are configured as HDFS data nodes and YARN node managers. 
Depending on the benchmark dataset size, as shown in Table~\ref{table:strategies}, 
each node manager is allocated a capacity of 8 CPU cores and 16 or 32 GB of memory.
SparkSQL jobs are submitted in the YARN client mode, 
running on 10 executors with 4 or 14 GB of memory and 1 or 2 CPU cores each. 
The distributed join parallelism is 20 and the Kyro library is used as the serializer for the shuffling
and broadcasting I/O. 
Unless otherwise specified, $w = 1$.

\textbf{Metrics}: The metrics that we use include the query completion time, optimization accuracy, 
and PSTS, which is a novel and general metric
for comparing the effectiveness of a join method selection strategy.  

\subsection{Query completion time}\label{section:queryTime}

We compare the query completion time of \methodName\ 
with those of various distributed join method selection strategies:
ShuffleSort, ShuffleHash, and Adaptive Query Execution (AQE).
The detailed descriptions of these strategies are presented in Table~\ref{table:strategies}. 
As broadcast NL join and Cartesian product join are ranked inferior to the strategies in the table, 
and they fail in most queries because of either out of memory or out of time, 
we do not evaluate these two join methods specifically.
ShuffleSort and ShuffleHash force the optimizer to select shuffle join methods 
using the join hint mechanism in SparkSQL, 
where users explicitly specify the preferred physical join methods in the query statement.
The AQE strategy~\cite{aqe18} is the most updated query optimization suite of SparkSQL 
and has been proven as robust for DISC platforms.
It selects the broadcast hash method instead of the shuffle methods 
if the absolute size of the adaptive runtime statistics of a joining dataset
is no larger than the user-defined size. 
The AQE optimization suite also includes other optimization features such coalescing small partitions and handling skewed shuffling. 
However, they are disabled in this evaluation to ensure a fair comparison of all strategies.
\methodName\ is similar to AQE in that it can select the broadcast hash method,
but it uses the relative cost model to judge whether broadcasting is appropriate.

\begin{figure}[!t]
	\centering
	\includegraphics[width = 1 \columnwidth]{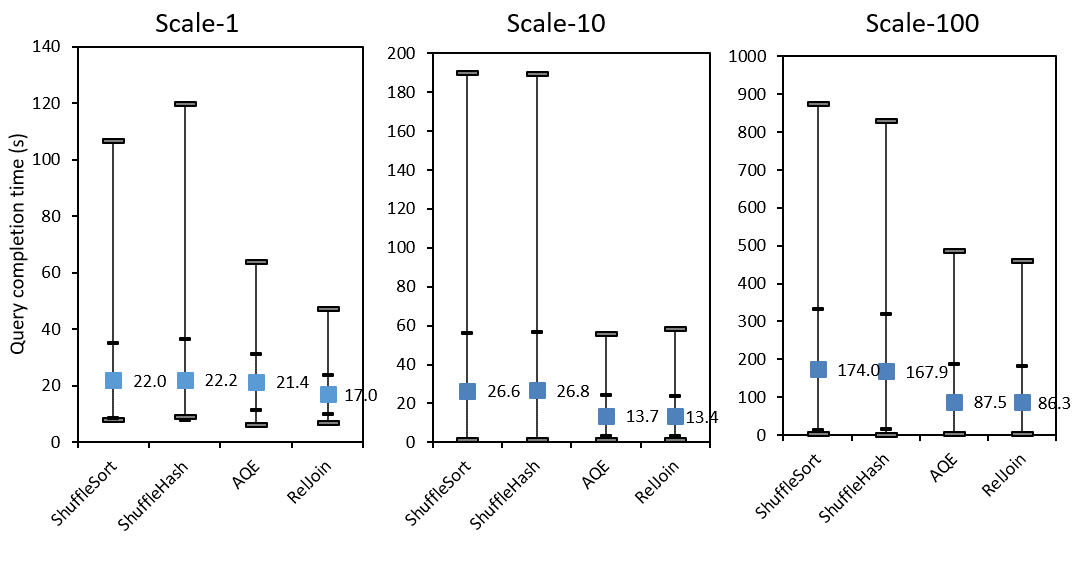}
	\caption{The average completion time of TPC-DS queries of various strategies in benchmarks of different scales}
	\label{fig:evalMethodTime}
\end{figure}

\methodName\ reduces the average and maximum query completion time significantly.
\methodName\ also reduces the standard deviation of the query time among different queries.
The query completion time results of the different strategies for benchmarks of different scales
are shown in Fig.~\ref{fig:evalMethodTime}.
The square blocks are the average time results, the higher and lower rectangle bars at both ends
are the maximum and minimum time results, and the smaller line bars are the average time $\pm$ standard deviation results. 
Compared to ShuffleSort, ShuffleHash, and AQE, 
\methodName\ reduces the average query time by 22.7\%, 23.6\%, and 20.8\%, respectively, in the scale-1 benchmark;
by 49.3\%, 49.8\%, and 2.0\%, respectively, in the scale-10 benchmark; 
and by 50.4\%, 48.6\%, and 1.3\%, respectively, in the scale-100 benchmark.
In addition, both AQE and \methodName\ significantly
decrease the maximum query time by allowing the broadcast hash join method. 
Compared to ShuffleSort, ShuffleHash, and AQE,
\methodName\ reduces the maximum query time by 55.6\%, 60.5\%, and 26.1\% in the scale-1 benchmark
and by 47.5\%, 44.6\%, and 5.5\% in the scale-100 benchmark, respectively. 
\methodName\ also has a lower standard deviation of query time than 
the other strategies, indicating that the query performance is less dispersed in \methodName.

We attributed the reduction in the average and maximum query time by AQE and \methodName\
compared with ShuffleSort and ShuffleHash
to the selection of the broadcast hash method when it is considered to be better than the other distributed join methods:
the shuffle sort and shuffle hash methods.
Broadcasting a (relatively) small dataset helps to decrease the network communication
workload by eliminating the need to shuffle the other large dataset. 
For example, query \textit{q72} in the scale-1 benchmark 
consists of ten logical joins, nine of which have massive left datasets
and small right datasets with a maximum size of 1.4 MB.
The completion time of query \textit{q72}, which is the maximum in both ShuffleSort (106 s)
and ShuffleHash (119 s), is reduced to 38 s for AQE and 26 s 
for \methodName. 
The broadcast hash join decreases the network traffic and the query time significantly. 

We further examine the optimization of another query in the scale-1 benchmark, 
\textit{q39b}, to determine how the cost model of \methodName\ makes a distributed join method selection. 
The completion times of query \textit{q39b} are 42 and 35 s
in ShuffleSort and ShuffleHash, respectively, but are reduced to 16 and 15 s
in BroadcastShuffle and \methodName, respectively. 
Query \textit{q39b} has seven logical joins and \methodName\ selects 
the broadcast hash method for six joins, 
in one of which the left and right datasets are approximately 40 and 0.13
MB, respectively.
According to Equations~\ref{eq:broadcastHashCost} and~\ref{eq:shuffleSortCost}, 
the costs of the broadcast hash join and the shuffle sort join are 45.2 MB and 78.4 MB, respectively.
By selecting the broadcast hash method,
\methodName\ outperform ShuffleSort and ShuffleHash
in the cases of joins with a relatively small dataset.

However, the absolute size criterion in AQE
may abuse the broadcast hash join method selection. 
We examine the join method selection results for all query plans in the scale-1 benchmark.
AQE and \methodName\ make 600 and 540 broadcast hash join selections, respectively, 
out of the 629 joins.
More broadcast hash joins than the appropriate 
results in AQE perform worse than \methodName\ in this benchmark. 
As the dataset scale increases, 
more datasets exceed the threshold for AQE to select the broadcast hash join method.
AQE decreases the number of broadcast hash join method selections, 
behaving more like \methodName\ and approaching its query performance. 
Despite this, \methodName\ retains the ability to select the broadcast hash join method
in larger-scale benchmarks to reduce the query time for queries in which both joining datasets 
have large absolute sizes, but one is relatively smaller than the other. 
This explains why AQE achieves similar average query time results to \methodName,
but \methodName\ can still decrease the maximum query time by 5.5\% in the scale-100 benchmark.

\begin{figure}[!t]
	\centering
	\includegraphics[width = 0.8 \columnwidth]{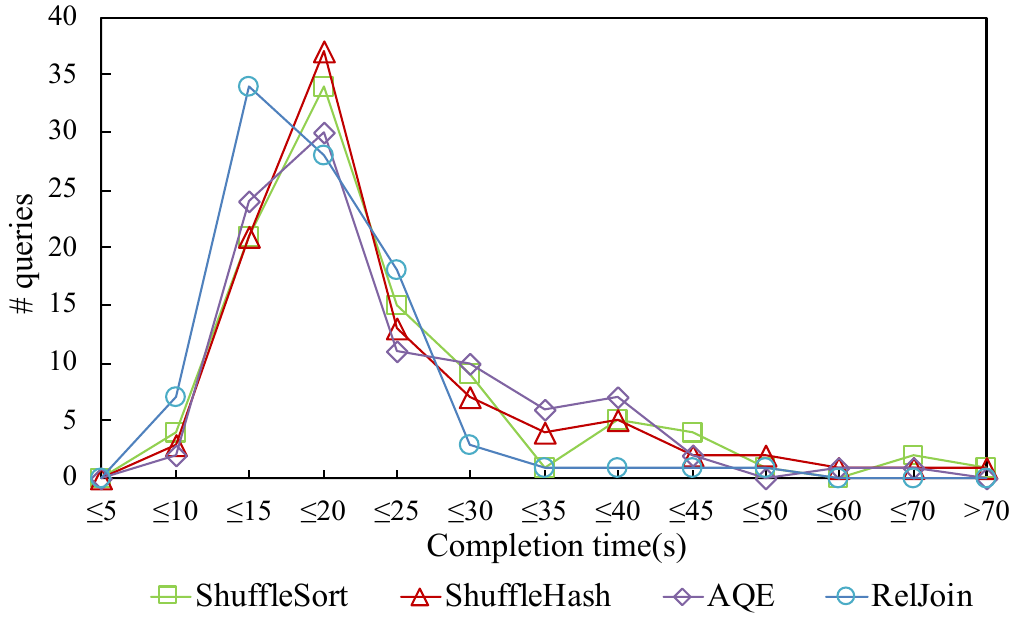}
	\caption{The distribution of query completion time of various strategies in the scale-1 benchmark}
	\label{fig:evalTimeDistribution}
\end{figure}

The distribution of the query completion time in the scale-1 benchmark in Fig.~\ref{fig:evalTimeDistribution} 
shows how \methodName\ improves the average query time in the bigger picture. 
The peak of \methodName\ in the query time distribution is to the left of those of the other strategies.
The numbers of queries that are completed within 15 s 
by ShuffleSort, ShuffleHash, AQE, and \methodName\
are 25, 24, 5, and 41, respectively. 
The numbers within 20 s are 59, 61, 35, and 69, respectively.
\methodName\ decreases the completion time for some queries known as outliers, 
as well as for a wide set of queries.
However, AQE has a fatter tail, indicating that 
it has more queries with a longer completion time. 
This is because it may select the broadcast hash join method 
when broadcasting introduces no performance benefit or, worse, costs more than shuffling. 
This is discussed in more details in Section~\ref{section:evalCriterion}.

\subsection{Optimization accuracy}\label{section:evalAccuracy}
When comparing a set of selection strategies, 
if the strategy that results in the lowest query completion time is viewed as optimal for a query,
the query-wise optimization accuracy of \methodName\ is as high as 63.9\%,
which is much higher than that of the other strategies.
Table~\ref{table:bestStrategy} presents the number of queries for which
each strategy is optimal 
and the corresponding optimization accuracy (percentage of optimized queries) 
for benchmarks of different scales. 
Without \methodName\ in the comparison, the optimization accuracy results 
of ShuffleSort and ShuffleHash are close to one another, whereas those of AQE are higher. 
A comparison of all four strategies shows that \methodName\ makes the best selection for 62 of the 97 queries,
45 of the 93 queries, and 36 of the 90 queries
in the scale-1, scale-10, and scale-100 benchmarks, respectively. 
As the queries in TPC-DS are complex queries with multiple joins, 
\methodName\ exhibits better results by selecting distributed join methods 
with the lowest cost for each logical join and generating the optimal
combination of a sequence of join methods. 
This is the major optimization idea of \methodName\, which distinguishes it from other strategies, 
and it outperforms them in most queries. 
As discussed in Section~\ref{section:queryTime}, 
with larger datasets in the TPC-DS benchmark,
AQE narrows the performance gap with \methodName\
by selecting fewer broadcast hash joins and behaves more like \methodName. 
Consequently, the optimization accuracies of these two strategies are similar for the scale-100 benchmark. 

\begin{table}[!t]
	\caption{The number of queries that a strategy achieves the least query time 
		and the corresponding optimization accuracy in benchmarks of different scales.
		The results with \methodName\ excluded (First-3) and included (All) 
		in the comparison are shown.  
		The results are represented as \#number(\%accuracy).}\label{table:bestStrategy}
	\begin{center}
		\begin{tabular}{ccccccc}
			\Xhline{2\arrayrulewidth}
			& \multicolumn{2}{c}{\textbf{Scale-1}}& \multicolumn{2}{c}{\textbf{Scale-10}} & \multicolumn{2}{c}{\textbf{Scale-100}}\\
			\cline{2-7}
			\textbf{Strategy}    & First-3 & All & First-3 & All & First-3 & All\\
			\Xhline{2\arrayrulewidth}
			ShuffleSort & 33(34.0) & 14(14.4) & 12(12.9) & 12(12.9) & 2(2.22) & 2(2.22)\\
			ShuffleHash & 28(28.9) & 9(9.3) & 7(7.5) & 7(7.5) & 10(11.1)& 10(11.1)\\
			AQE & 36(37.1) & 12(12.4) & 74(79.6) & 32(34.4) & 78(86.7) & \textbf{42(46.7)}\\
			\methodName & - & \textbf{62(63.9)} & - & \textbf{45(45.2)}& - & \textbf{36(40.0)} \\
			\hline
			Total & 97(100) & 97(100) & 93(100) & 93(100) & 90(100) & 90(100)\\
			\Xhline{2\arrayrulewidth}
		\end{tabular}
	\end{center}
\end{table}

Different local join methods that are applied by distributed joins are minor  
factors influencing the query performance. 
Fig.~\ref{fig:evalMethodTime}, Fig.\ref{fig:evalTimeDistribution} 
, and Table~\ref{table:bestStrategy} all indicate that 
the shuffle sort and shuffle hash methods have very similar results
in terms of the average query time, query time distribution,
and optimization accuracy. 

\subsection{Relative size vs. absolute size: criterion for broadcasting}\label{section:evalCriterion}
Hereafter, we use the results of the scale-1 benchmark to explore additional properties 
and mechanisms of \methodName. 

We further claim that the relative size of the two joining datasets used by \methodName\
is a better criterion than the absolute size when using the broadcast hash method. 
AQE and \methodName\ are similar strategies that allow broadcast
hash joins when they judge a dataset to be sufficiently small for broadcasting. 
They differ in the criterion of determining sufficiently small datasets. 
The former uses the user-configured absolute size, whereas the latter uses the relative size
considering the distributed join parallelism and network cost weight. 

We delve deeper into the query plans and join method selection results for these
two strategies to verify our claim.
As our focus is the performance difference between broadcasting and shuffling,
and the shuffle sort and shuffle hash methods have similar performances,
we consider these shuffle methods as the same method.
We sort out the distributed join methods and corresponding statistics, 
where one strategy selects the broadcast hash method and the other
selects a shuffle method.
For a query in which the strategies make different join method selections, 
the cost difference based on our cost model and the query completion time difference are recorded. 
Among the 629 selections in the scale-1 benchmark, 
these two strategies make different selections in 66 cases, 
in 63 of which \methodName\ selects shuffle methods 
and AQE selects broadcast hash methods, 
and in three of which they do so in reverse. 
The benchmark cost difference is 5834.2 MB,
and the benchmark completion time difference is 419.9 s
which is 20.8\% of the AQE benchmark completion time, 
as indicated in Table~\ref{table:planStat}.
With everything else remaining the same, the performance improvement is attributed to the 
join method selection difference.
Each different method selection made by \methodName\
contributes to a cost reduction of 88.4 MB and a query completion time
reduction of 6.4 s.

\begin{table}[!t]
	\caption{Comparing \methodName\ with AQE, 
		the statistics of (1) the number of different and total join method selections, (2) the resulted cost difference, 
		(3) the resulted query time reduction, (4) the cost difference per number of different join method selections, 
		(5) the query time difference per number of different join method selections, 
		(6) the percentage of different join method selections to the total number of joins, 
		(7) the percentage of time reduction, 
		and (8) the performance sensitivity to selections (PSTS).}
	\label{table:planStat}
	\begin{center}
		
		\begin{tabular}{lc}
			\Xhline{2\arrayrulewidth}
			\textbf{Statistics} & \textbf{Value} \\
			\Xhline{2\arrayrulewidth}
			(1) \#JoinDiff (\#Join) & 66 (629)\\
			(2) CostDiff & 5834.2 MB \\
			(3) TimeDiff & 419.9s \\
			(4) CostDiff / \#JoinDiff & 88.4 MB\\
			(5) TimeDiff / \#JoinDiff & 6.4s \\
			(6) \%JoinDiff & 10.5\% \\
			(7) \%TimeDiff & 20.8\% \\
			(8) \%TimeDiff / \%JoinDiff (\textbf{PSTS}) & \textbf{1.98}\\
			\Xhline{2\arrayrulewidth}
		\end{tabular}
	\end{center}
\end{table}

However, the above metrics rely on benchmark-specific properties, such as the size of the dataset
and number of queries, and are not comparable across benchmarks to 
reflect the optimization effectiveness of different strategies directly. 
We propose the term \textit{Performance Sensitivity To Selections} (PSTS)
as a metric to measure the query optimization effectiveness of a join method selection strategy.
The PSTS is calculated as the percentage of the time difference divided by
the percentage of the join method selection difference, which indicates the extent to which the 
query completion time is affected by the optimization result of a selection strategy
with AQE as the baseline.
A PSTS of 1 indicates that a 1\% difference in the number of join methods
selected by a selection strategy
contributes to a 1\% reduction in the benchmark completion time.
In general, a higher PSTS indicates that the query optimization is more effective .
When the PSTS is close to 0 or even negative, 
the optimization of a selection strategy is ineffective
or even negatively affects the query performance.
In the TPC-DS case, the PSTS of \methodName\ is 1.98, which suggests that
\methodName\ effectively
optimizes the query plan compared with AQE. 
As a reference, the PSTSs of ShuffleSort and ShuffleHash in TPC-DS are $-0.03$ and $-0.04$, respectively,
which indicates that they have slightly negative optimization effects compared with AQE.
The results of the PSTS do not include benchmark-specific properties and are comparable across benchmarks
for optimization effectiveness. 

\methodName\ outperforms AQE mainly by not using the broadcast hash method
when one dataset is not sufficiently smaller than the other in a join.
Among the 66 cases of inconsistent join method selections, 
\methodName\ resolves that shuffle methods are better selections in 63.
\methodName\ selects the broadcast hash method only when 
the larger dataset is 39 times as large as the smaller dataset ($k_0 = 39$). 
However, AQE selects the broadcast hash method as 1.10 MB is smaller 
than the broadcast size threshold (10 MB). 
When considering the selection of a broadcast hash method, 
the relative size criterion used by \methodName\ is better than the absolute size criterion
used by AQE. 



\begin{figure}[!t]
	\centering
	\includegraphics[width = 0.8 \columnwidth]{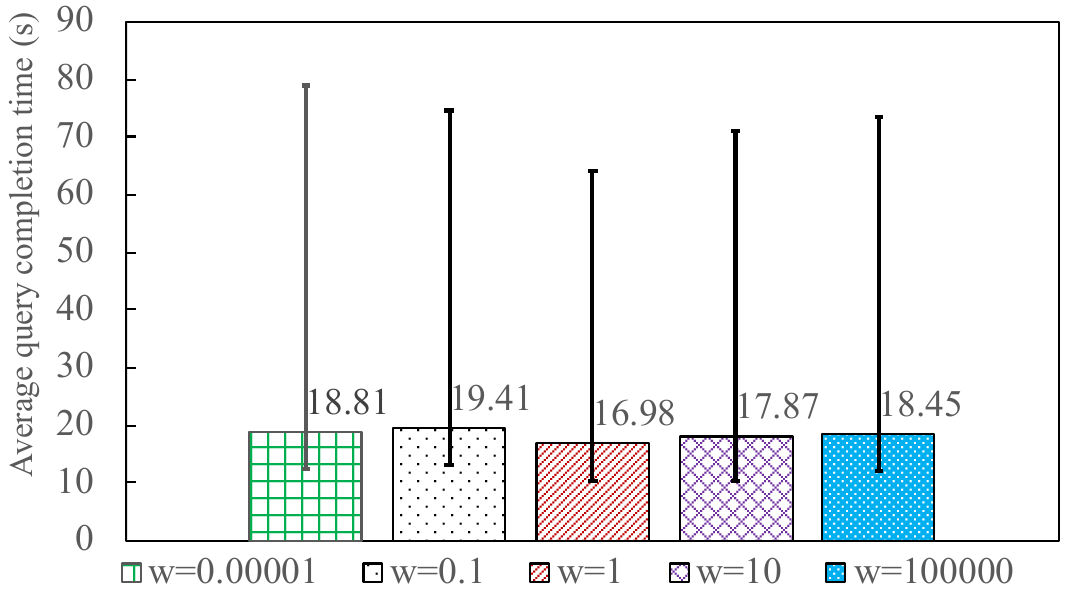}
	\caption{The average query completion time of \methodName\ with different $w$ values in the scale-1 benchmark}
	\label{fig:evalWTime}
\end{figure}

\subsection{The impact of $w$}\label{section:evalWImpact}
We explore the impact of different values of $w$, which is
the relative weight of the network cost on the computing cost.
Fig.~\ref{fig:evalWTime} shows that \methodName\ has similar average, maximum, and minimum query times 
in the scale-1 TPC-DS benchmark with $w$ at different scales.

Different configurations of $w$ have a relatively higher impact on the maximum query time
than on the average query time. 
The trend of the maximum query time exhibits an obvious ``V'' shape, with $w=1$ at the bottom of the valley, 
which is 21.4\% and 14.4\% lower than that when $w=0.00001$ and $w=100,000$, respectively.
This suggests that the optimal configuration of $w$ is approximately 1 in our testbed. 
When $w$ is substantially larger than 1 (for instance, as large as 100,000), the overall
cost is determined entirely by the network cost.
The average query time of 18.45 s is slightly larger than that when 
$w = 1$, but is still more than 16\% smaller than the query time of 
the ShuffleSort, ShuffleHash, and AQE strategies (Fig.~\ref{fig:evalMethodTime}).
We also test the case of $w = 0.1$, where the overall cost is mainly determined by the local computing cost, 
which may occur in the fast network, e.g., the zero-copy networking of the Remote Direct Memory Access (RDMA) technology, and slow memory scenarios.
Owing to the build phase, which builds a hash map for the entire replicate dataset in each task, 
the local computing cost of the broadcast hash method is higher than those of the shuffle methods.
\methodName\ behaves more similarly to ShuffleSort and ShuffleHash, 
and tends to select shuffle methods when shuffling network traffic 
is not significant compared with broadcasting. 
The average and maximum query times both increase by approximately 15\% compared to 
the case of $w=1$. 
Weakening or overemphasizing the network cost
slightly degrades the query performance.
Regarding the ``V'' trend, tuning the optimal $w$ is similar to a convex problem,
where users can start with an 
empirically near-optimal value first, such as $w=1$, and attempt values leading to the downtrend.  

We have highlighted some important evaluation results. 
First, compared with AQE, \methodName\ reduces the average and maximum query completion times
by 20.8\% and 26.1\%, respectively.
Second, \methodName\ achieves the best completion time for 62 of the 97 queries, 
which is five times as high as that of AQE. 
Third, with AQE as the baseline, the PSTS of \methodName\ is 1.98 for the scale-1 benchmark, 
which indicates that 1\% of the join selection optimization contributes to a 1.98\% reduction in
the query completion time. 
Fourth, tuning the hyperparameter $w$ is straightforward. 
Setting $w$ either too large or too small will slightly degrade the performance, 
but it still performs better than AQE.

\section{Related work}\label{section:relatedWork}
Cost-based join query optimization has long been an important topic 
in database query optimization. 
It often encounters the combination explosion problem,
resulting in a locally optimal plan. 
Oracle~\cite{costOracle06, subqueryOpt09, queryOpt15} uses a set of rules 
to generate many combinations of physical alternatives
for a sequence of logical operations, following which it uses
cost-based optimization to search for the optimal query plan with the lowest cost,
which can easily run into potential combination explosion problems.
To address this problem, Oracle performs a quasi-random walk 
in the search, which achieves a local minimum cost at every step. 
\methodName\ also performs a quasi-random walk to select the least-cost distributed join method
for a sequence of logical join operations in the query plan. 
However, for the optimization problem of distributed join method selection, 
because the cost and selection of a distributed join method for a logical join 
are independent of those for other logical joins in the query plan, 
the best join method for each logical join leads to 
the globally optimal physical plan.

Most cost-based join optimizations~\cite{joinOrder09,queryOptGood15,costAdaptive09,randomJoinOrder97,joinOrder03} 
work on cost-efficient methods to achieve the optimal join order, 
by using heuristic, randomized, or genetic algorithms.
The accuracy of dataset statistics is crucial for determining the optimal join order. 
Some studies have obtained statistics by sampling~\cite{randomSampleStat90} 
or adapted statistics based on the runtime statistics 
from every execution stage~\cite{adaptiveBig13,adaptiveStat17}. 
In recent years, many machine-learning-based models~\cite{costOptAI14, learnedCost19,deepCardinality, learnedRewrite21}
have been developed to estimate the dataset cardinality and to rewrite the query plan for the optimal join order.
\methodName\ can work together with join order optimization strategies to
achieve the best performance.
It uses adaptive runtime statistics to select the best physical join method for logical joins that are already
in the optimal join order.   

Many cost models for join methods focus on the local join cost for query optimization 
and do not precisely incorporate the network cost of the distributed environment. 
Some models~\cite{sortVsHash13, sortVsHash09,memoryJoin02, memoryCost02} 
analyze the complexity of the algorithm implementation of the hash join and sort join,
and compare their costs based on the use of CPU, memory, and disk I/O in different computing phases. 
Other methods~\cite{forecastCost15,predictTime13} predict the query time of joins 
by calibrating the cardinality workloads, measuring the speeds of local computing resources, 
and analyzing the memory I/O access patterns. 
Some methods~\cite{rackRdmaJoin15,distributedJoin17} model the partitioning cost 
of shuffle-like distributed joins via RDMA transfers in the data exchange phase, 
and mainly compare the cost differences in the local join phase,
but not the network costs of different data exchange methods. 
\methodName\ explores and compares the broadcast and shuffle costs in the data exchange
phase of various distribution methods and emphasizes the network cost in the cost model.

\begin{table}[!t]
	\caption{A summary of the comparison of the characteristics of \methodName\ with 
	other distributed join cost models}
	\label{table:relatedWork}
	\begin{center}
		
		\begin{tabular}{l|ccccccc|c}
			\Xhline{2\arrayrulewidth}
			 &\cite{forecastCost15}&\cite{distributedJoin17}&\cite{distanceJoin23} &\cite{costSpark18} & \cite{costSparkSqlJoin18}  & \cite{skewJoinSpark22} & \cite{compJoinSpark21} & Ours \\
			\Xhline{2\arrayrulewidth}
			Workload-based &$\checkmark$ & $\checkmark$ & $\checkmark$ & & $\checkmark$ & $\checkmark$ & $\checkmark$ & $\checkmark$\\
			Time-based & & & & $\checkmark$ & & & &\\
			\hline
			Rely on data volume &$\checkmark$ & $\checkmark$& $\checkmark$ &$\checkmark$ & $\checkmark$& $\checkmark$& $\checkmark$ &$\checkmark$\\
			Rely on access patterns & $\checkmark$ & & & & & & & \\
			Rely on I/O throughput &  & $\checkmark$ & $\checkmark$ &$\checkmark$ & & $\checkmark$& $\checkmark$ &\\		
			Require abundant tuning& & & & & $\checkmark$ & & & \\ 
			\hline
			Consider resource allocation & & & &$\checkmark$ & & & $\checkmark$ & $\checkmark$\\
			Consider data relations & & $\checkmark$& $\checkmark$& $\checkmark$ & &$\checkmark$&$\checkmark$& $\checkmark$\\
			Robust with data skew & & & $\checkmark$& & & $\checkmark$ & & $\checkmark$\\ 
			\Xhline{2\arrayrulewidth}
		\end{tabular}
	\end{center}
\end{table}

Many cost models~\cite{costSpark18,distanceJoin23,costSparkSqlJoin18,skewJoinSpark22,compJoinSpark21} also consider the network cost of distributed
join methods; however, they are highly dependent on the hardware access patterns whose performance values are 
difficult to determine on scalable platforms or involve abundant
hyperparameters that are specific to the complex distributed environment, 
leading to potential practical problems.
For example, the cost model proposed by Baldacci and Golfarelli~\cite{costSpark18} 
requires accurate measurement of the disk read/write throughput between processes
and the network throughput between nodes in the same rack or across racks. 
The cost model proposed by Lian and Zhang~\cite{costSparkSqlJoin18} requires the tuning of several hyperparameters matrices 
based on the empirical correlative cost of all physical plan candidates. 
There are also concerns regarding data skew. 
Phan et al.~\cite{compJoinSpark21} modeled the costs of various distributed join methods 
but provided no evidence of robustness in data skew cases. 
Some cost models~\cite{distanceJoin23,skewJoinSpark22} address the data skew problem 
by incorporating the skewness assumption into the problem setting.
However, the above cost models fail to realize the importance of the relative size of the joining datasets 
in determining the join costs. 
\methodName\ relates the cluster workload to the relative sizes of the joining datasets
and has only one hyperparameter, which does not sensitively affect the query performance
in modern network environments.
Moreover, our analysis of the impact of data skew showed that the cluster workload modeling method 
makes \methodName\ robust even when data skew exists. 
Table~\ref{section:relatedWork} compares the characteristics of 
\methodName\ with those of other distributed join models.

Other optimizations of distributed joins~\cite{parallelJoinDistributed17,confluence17,broadcastJoin14} 
include considering the relationship and locality of the data of multiple join operations,
such that data can be joined locally with less network traffic  
during a sequence of joins. 
These optimizations can help to refine the data exchange phase cost and performance 
in \methodName.

\section{Conclusions and future work}\label{section:conclusion}
In this study, we have developed and compared the cost models of various distributed join methods
that precisely estimate network workloads using adaptive runtime statistics. 
We integrated the distributed join cost model into \methodName, 
which is an efficient query optimization strategy that selects the best distributed join methods
for logical joins in the query plan. 
The evaluation results for the TPC-DS benchmark show that \methodName\
is optimal for most queries 
and significantly improves the query completion time.

In the future, we will work on learning-based cost models to
connect workloads to the query completion time 
such that minimizing the query cost is directly related to the target of 
minimizing the query completion time.
Moreover, the cost-based distributed join method selection strategy in query plan optimization
motivates us to discover more potential variants of distributed join implementations
that can optimize the join query performance in specific cluster environments with specific query workloads. 




\bibliographystyle{elsarticle-harv}
\bibliography{relJoin-major}


%
%
%
\end{document}